\documentclass[%
 aip, reprint,
rsi,
 amsmath,amssymb,
 Conference Proceedings
]{revtex4-1}

\usepackage{graphicx}% Include figure files
\usepackage{dcolumn}
\usepackage{bm}% bold math

\usepackage{xcolor}
\usepackage[utf8]{inputenc}
\usepackage[T1]{fontenc}
\usepackage{mathptmx}
\usepackage{etoolbox}
\usepackage{amsfonts,amsmath,amssymb}
\usepackage{multirow}
\usepackage{hyperref}% add hypertext capabilities
\usepackage[normalem]{ulem}
\hypersetup{colorlinks=true,linkcolor=cyan,citecolor=cyan,filecolor=cyan,urlcolor=cyan}

\usepackage{orcidlink}

\makeatletter
\def\@email#1#2{%
 \endgroup
 \patchcmd{\titleblock@produce}
  {\frontmatter@RRAPformat}
  {\frontmatter@RRAPformat{\produce@RRAP{*#1\href{mailto:#2}{#2}}}\frontmatter@RRAPformat}
  {}{}
}%
\makeatother
\begin{document}

\preprint{AIP/123-QED}

\title[Thomson Scattering diagnostic for SMART]{Design of a Thomson scattering diagnostic for the SMART tokamak}
% Force line breaks with \\
\author{M. Kaur}
 %\orcidlink{0000-0001-6008-6676}
 \email{mkaur@pppl.gov}
\author{A. Diallo}%
\author{B. LeBlanc} 
%\author{J. Clark}
\affiliation{ Princeton Plasma Physics Laboratory, Princeton, NJ 08540, USA.}%
\author{J. Segado-Fernandez}
\author{E. Viezzer}
\affiliation{Department of Atomic, Molecular and Nuclear Physics, University of Seville, Seville, Spain.}
\author{Roger Huxford }
\affiliation{RBH Optics, Burgess Hill, West Sussex RH15 8HL, United Kingdom}
\author{A. Mancini}
\author{D. J. Cruz-Zabala}
\affiliation{Department of Atomic, Molecular and Nuclear Physics, University of Seville, Seville, Spain.}
\author{M. Podesta}
\affiliation{ Princeton Plasma Physics Laboratory, Princeton, NJ 08540, USA.}
 \affiliation{Presently at Ecole Polytechnique Fédérale de Lausanne, Swiss Plasma Center, CH-1015 Lausanne, Switzerland.}
 %\verb|\footnotemark|\footnotemark{}
\author{J. W. Berkery}
\affiliation{ Princeton Plasma Physics Laboratory, Princeton, NJ 08540, USA.}
\author{M. Garcia-Muñoz}
\affiliation{Department of Atomic, Molecular and Nuclear Physics, University of Seville, Seville, Spain.}

\vspace{-2mm}
\begin{abstract}

We describe the design of a Thomson scattering (TS) diagnostic to be used on the SMall Aspect Ratio Tokamak (SMART). SMART is a spherical tokamak being commissioned in Spain that aims to explore positive triangularity (PT) and negative triangularity (NT) plasma scenarios at a low aspect ratio. 
The SMART TS diagnostic is designed to enable a wide range of electron temperature ($1$ eV to $1$ keV) and density ($5 \times 10^{18}$ to $1 \times 10^{20}~\rm{m^{ - 3}}$) measurements. A $2$ Joule laser operating at $1064$ nm will be used to probe the electron temperature and density of the plasma. The laser is capable of operating in the burst mode at $1$ kHz, $2$ kHz, and $4$ kHz to investigate fast phenomena or at $30$ Hz to study $1$ sec (or more) long discharges. 
The scattered light will be collected over an angular range of $60^{\circ}$ – $120^{\circ}$ at 28 spatial points in the midplane covering the core region and edge plasma on both the low-field side (LFS) and the high-field side (HFS). Simulation data is used to determine the optimum location of Thomson scattering measurement points to effectively resolve the edge pedestal in the LFS and HFS regions under different triangularity conditions. 
Each scattering signal will be spectrally resolved on five wavelength channels of a polychromator to obtain the electron temperature measurement. We will also present a method to monitor \textit{in-situ} laser alignment in the core during calibrations and plasma operations.

\end{abstract}

%\vspace{-2mm}

\maketitle
\section{\label{sec:level1}Introduction}
%\vspace{-4mm}
The production of high density plasmas without triggering magnetohydrodynamic instabilities like Edge Localized Modes (ELMs) is an important requirement for realizing magnetic confinement fusion in a tokamak.  
Among the various tokamak configurations, spherical tokamaks (ST) offer compact-configuration plasmas. 
Moreover, negative-triangularity (NT)-shaped tokamak plasmas offer a wide range of advantages such as high-energy confinement, non-ELMing plasmas, and detached-divertors along with high-density operations \cite{camenen2007impact,austin2019achievement}. Here triangularity ($\delta$) refers to the shape of the poloidal cross-section of the last closed flux surface or the separatrix of a tokamak. To combine the benefits of spherical tokamaks and negative-triangularity, the SMART tokamak is currently being commissioned at the University of Seville in Spain that aims to explore the physics of NT plasmas and compare it with PT counterpart scenarios\cite{Mancini_2023, segado_2023}.

We describe the design of the TS diagnostic for SMART to measure the spatially and temporally resolved electron temperature ($T_{\rm{e}}$) and density ($n_{\rm{e}}$)\cite{Kaur_2023}. A TS diagnostic works by accelerating free plasma electrons using a linearly-polarized short-pulse laser beam. 
Accelerated electrons emit electromagnetic radiation. 
The scattered radiation is called TS radiation, which is collected and resolved spectrally on different wavelength bands to obtain $T_{\rm{e}}$ from spectral broadening and $n_{\rm{e}}$ from the amplitude\cite{prunty2014primer}. Knowledge of $T_{\rm{e}}$ and $n_{\rm{e}}$ allows one to understand the effects of various plasma heating and confinement techniques on tokamak plasmas and the underlying physics in different regimes, for example, electron heat transport, particle transport, etc. 
The primary goal of the TS diagnostic on SMART is to facilitate physics studies by precisely resolving the core and the edge (pedestal) for both NT and PT configurations. This diagnostic provides detailed measurements across the HFS and LFS regions of the magnetic axis of SMART, as well as in the core plasma.

The rest of the article is arranged as follows. In section~\ref{Sec:Overview}, we provide an overview of the SMART tokamak and briefly describe the operational requirement of the TS diagnostic. The design of the TS diagnostic along with all its major components is given in section~\ref{Sec:TS_diagnostics}, followed by a summary in section~\ref{Sec:summary}. 

\section{Overview }\label{Sec:Overview}

\begin{figure}
	\includegraphics[scale=0.5]{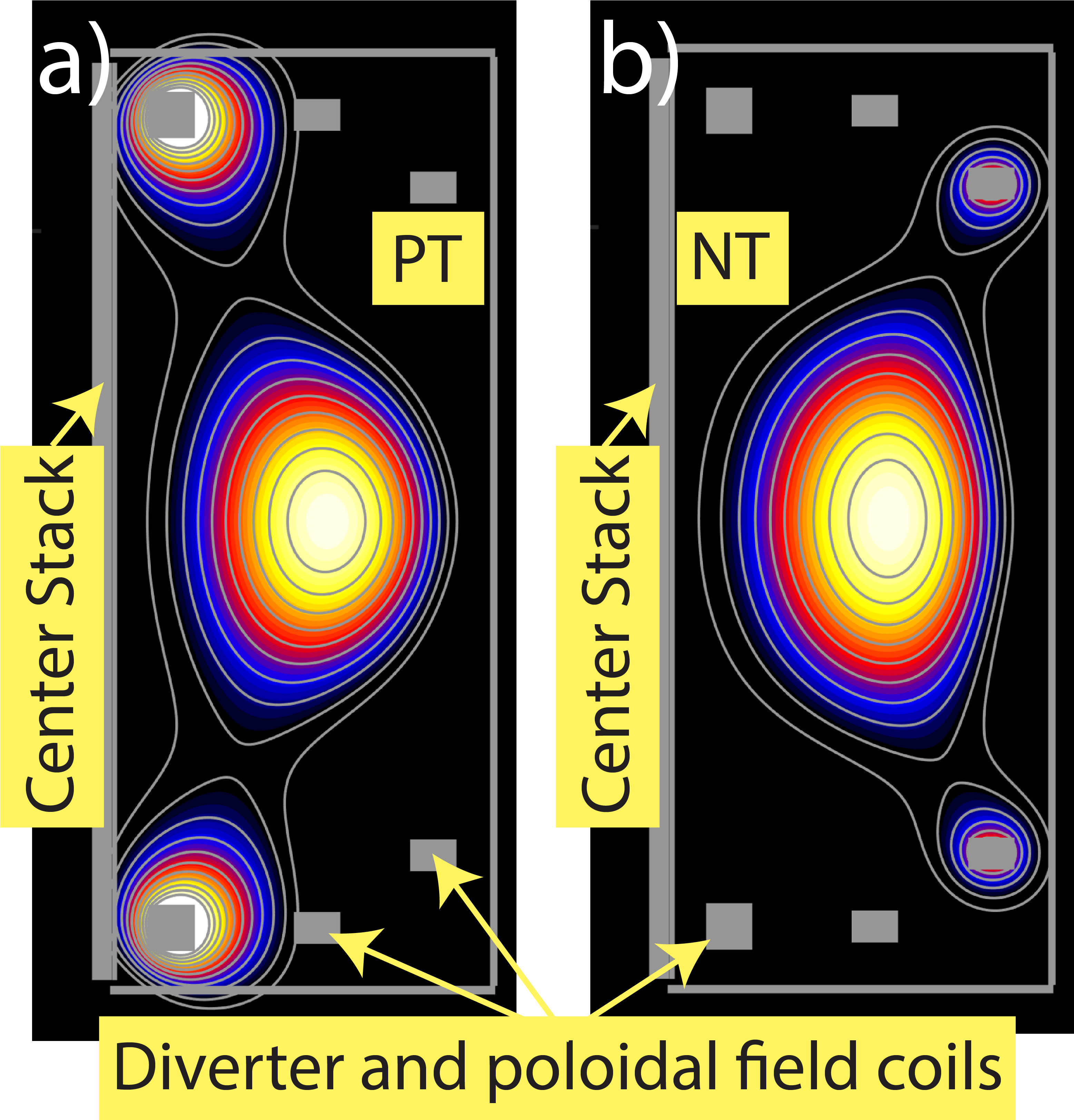}
	\caption{\label{fig:smart} a) The PT and b) NT plasma scenarios from Cruz-Zabala \textit{et al.}\cite{Zabala} are shown in a poloidal cross-section of the SMART tokamak along with the vessel wall and in-vessel diverter and poloidal field coils. }
\end{figure}

SMART is designed for physics studies by shaping plasma with $\delta$ ranging from $-0.6$ to $+0.6$, and the aspect ratio can be varied from $1.4$ to $3$ \cite{Mancini_2023}. A poloidal cross-section of the SMART tokamak overlaid with PT and NT configurations is shown in Fig.~\ref{fig:smart} a) and b), respectively. 
The operation of SMART is foreseen in three phases, see Table~\ref{table:SMART_phases}. The toroidal magnetic field ($B_{\rm{t}}$) will be varied from $0.1$~T in Phase 1 to $1$~T in Phase 3, and the plasma current ($I_{\rm{p}}$) ranges from $100$~kA in Phase 1 to $1$~MA in Phase 3. The SMART plasmas will be heated in Phases 2 and 3 with neutral beam injection (NBI) to attain high temperatures. 

\begin{table}[h!]
 \begin{tabular}{ |p{1.8cm}|p{1.8cm}|p{1.8cm}|p{1.8cm}|  }
 \hline
 Parameters & Phase 1 & Phase 2 & Phase 3 \\
 \hline
 $I_{\rm{p}}~[\rm{kA}]$ & $100$  & $200 - 500$  & $>500$  \\
 \hline
 $B_{\rm{t}}~[\rm{T}]$ & 0.1 & 0.4 & 1.0 \\
 \hline
 $\tau_{\rm{pulse}}~[\rm{sec}]$ & 0.15 & 0.5 & >1 \\
 \hline 
 $P_{\rm{ECH}}~[\rm{kW}]$ & 6 & 6 & 200 \\
 \hline
 $P_{\rm{NBI}}~[\rm{MW}]$ & - & 0.3 - 1 & 1 \\
 \hline
\end{tabular} 
\caption{Table of operating parameters in different SMART phases\cite{Mancini_2023, segado_2023, Podesta_2024}. Here $P_{ECH}$ stands for electron cyclotron heating. }
\label{table:SMART_phases}
\end{table}

Based on the operational parameters given in Table~\ref{table:SMART_phases}, a literature survey of STs is conducted to estimate the $T_{\rm{e}}$ range of SMART plasmas in different phases of operation. The expected $T_{\rm{e}}$ range strongly influences the design of the diagnostic for a typical density, 
$n_{\rm{e}} \sim 10^{19}~\rm{m^{-3}}$. For example, the values of $T_{\rm{e}}$ determine the wavelength ranges of the TS radiation that need to be measured.  The survey to estimate $T_{\rm{e}}$ is summarized in Table~\ref{table:comparison}.

A comparison of Table~\ref{table:SMART_phases} and \ref{table:comparison} shows that in Phase 1, SMART is expected to produce a minimum core $T_{\rm{e}}$ of up to $100$~eV based on results from VEST/VEST-II \cite{Kim_2019, Kim_2021, Jung_2022} under similar $I_{\rm{p}}$ and $B_{\rm{t}}$ ranges. Globus-M\cite{Gusev_2015} and START\cite{Sykes_1997} results suggest core $T_{\rm{e}}$ in the range from $350$~eV to $600$~eV in Phase 2 in the presence of an external heating source and higher $I_{\rm{p}}$ and $B_{\rm{t}}$. In Phase 3, SMART is anticipated to achieve $T_{\rm{e}}$ up to $1$~keV based on data from Globus-M2 \cite{Telnova_2021, Bakharev_2020, Kurskiev_2022} and ST40\cite{Gryaznevich_2022}. 
In the scrape-off layer (SOL), $T_{\rm{e}}$ may drop to a value as low as $1$ eV. Therefore, the SMART TS diagnostic is optimized for a wide range of expected SMART parameters under different operational phases to measure a wide $T_{\rm{e}}$ range from $1$~eV to $1$~keV. 

\begin{table}[h!]
 \begin{tabular}
 { |p{1.3cm}|p{1.2cm}|p{0.9cm}|p{1.3cm}|p{0.9 cm}|p{0.9 cm}|p{0.9 cm}|  }
 \hline
 Parameters & VEST/ VEST-II & START & Globus-M & \multicolumn{2}{|c|}{Globus-M2} & ST40 \\
 \hline
 $I_{\rm{p}} [\rm{kA}]$ & $50 - 150$  & $200$  & $120 - 250$ & $200-300$ & $350$ & $500 - 580$  \\
 \hline
 $B_{\rm{t}} [\rm{T}]$ & 0.15 & 0.3 & 0.4 & 0.7 & 0.9 & <1.0 \\
 \hline
 External Heating [MW] & - & 0.5 (NBI) & 0.1 (RF power) & 0.85 (NBI) & 0.7 (two D2 NBIs) & 1.8 \\
 \hline
 $T_{\rm{e}} [\rm{keV}]$ & 0.1 & 0.35 & 0.55 & 0.2 - 0.9 & 1.5 & 0.9 \\
 \hline
\end{tabular} 
\caption{Table of achieved core $T_{\rm{e}}$ in comparable STs at operating parameter values similar to different SMART phases. Please note that the operation windows of comparable STs beyond the planned operation phases of SMART are not considered for comparison.   }
\label{table:comparison}
\end{table}

\section{The SMART TS diagnostic}\label{Sec:TS_diagnostics}

The TS diagnostic is designed to cover the first three operational phases of SMART, featuring multiple subsystems that enable the measurements of $T_{\rm{e}}$ and $n_{\rm{e}}$ at various spatial locations and times during a plasma discharge. These subsystems are detailed in this section.

\subsection{High-repetition rate laser and beam transport}
 
An InnoLas-made Nd:YAG laser operating with $2$ Joule nominal energy per pulse at $30$ Hz and pulse width of $10$ ns at the fundamental wavelength of $1064$ nm is purchased.  
The beam divergence and the pointing stability of this laser are $11.68$ $\mu$rad and $0.5$ mrad (full angle), respectively.
This laser is capable of operating in a burst mode at reduced energy per pulse to achieve measurements at a high-repetition rate during a discharge. Here, a burst represents a group of pulses ($4$ to $6$ in our case) fired at a fast rate.   
The frequency of pulses in a burst can be set at $1$ kHz, $2$ kHz or $4$ kHz. For four pulses in a burst, the laser delivers $1.88$ J energy per pulse at $4$ kHz, $1.7$ J at $2$ kHz, and $1.57$ J at $1$ kHz burst. The energy per pulse drops further for $6$ number of pulses in a burst. Moreover, the burst can be repeated at every $\sim 200$~ms during a shot. 
This mode will be beneficial for exploring several fast phenomena and short-pulse discharges, for example, Phase 1 of SMART.

\begin{figure}
	\includegraphics[scale=0.45]{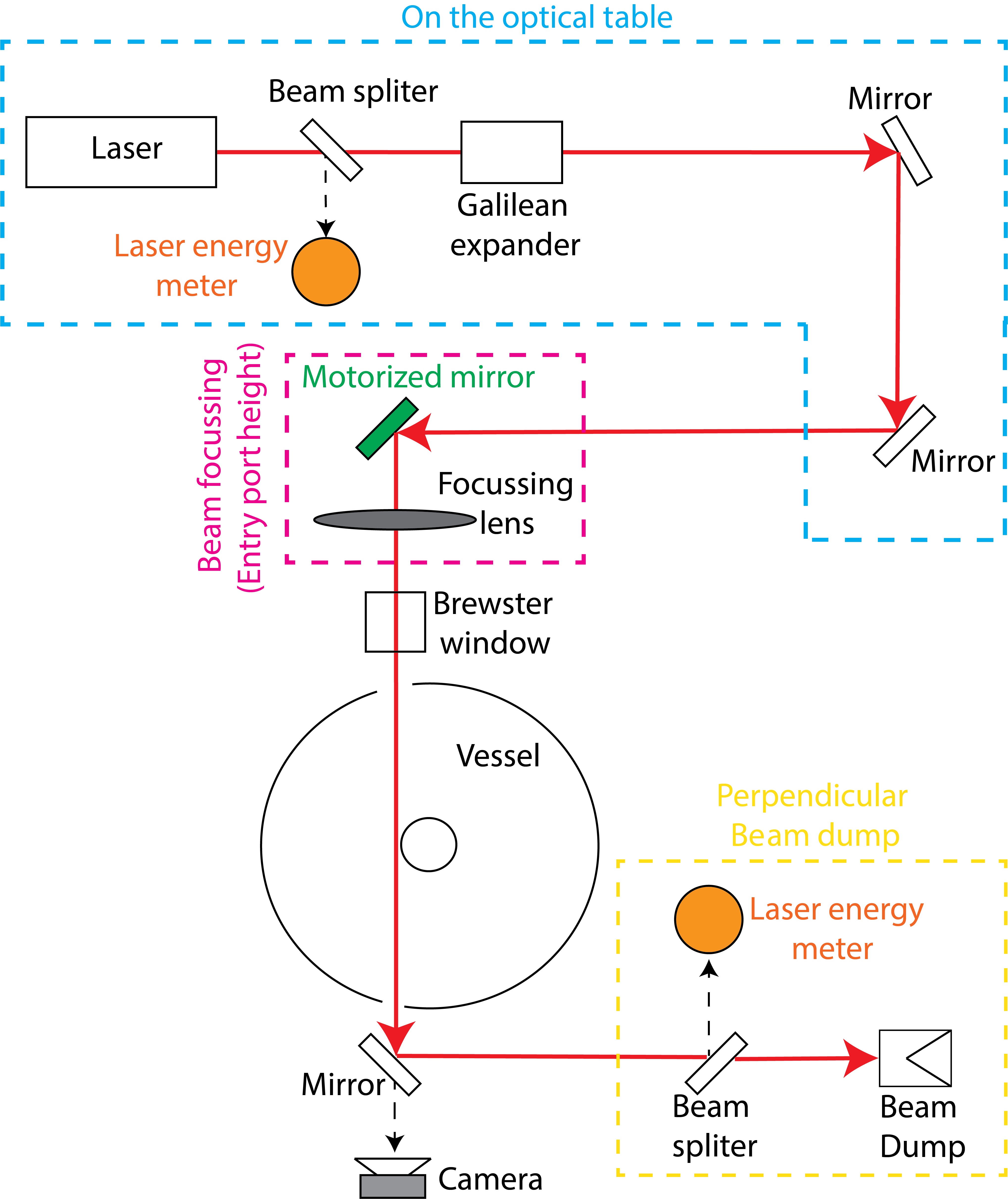}
	\caption{\label{fig:beam_transport} Here is a schematic of the laser beam transport and dump.}
\end{figure}

The laser and ancillary instruments will be located in a room $10$~m away from the SMART test cell. A schematic of the laser beam transport is shown in Fig.~\ref{fig:beam_transport}. Each laser pulse energy will be measured using a fast-response energy meter (repetition rate of 25 kHz) installed at the laser head. Upon exiting from the laser head, the laser beam will be expanded from $12$~mm diameter to $25$~mm using a Galilean beam expander to reduce beam divergence and to keep the energy load on to the entry window below the damage threshold. After that, a set of mirrors and lenses will focus the vertically polarized laser beam into the midplane through a Brewster window to make $T_{\rm{e}}$ and $n_{\rm{e}}$ measurements. Upon its exit from the main vessel, the laser beam energy will be measured, and the beam will be dumped on a Kentek-made beam dump housed outside the vacuum vessel. The laser is equipped with a visible alignment laser that will be used along with a camera for %coarse 
alignment of the laser prior to calibrations and operations. 

\subsection{TS scattering-location selection and collection optics}

\begin{figure*}
\includegraphics[scale=0.8]{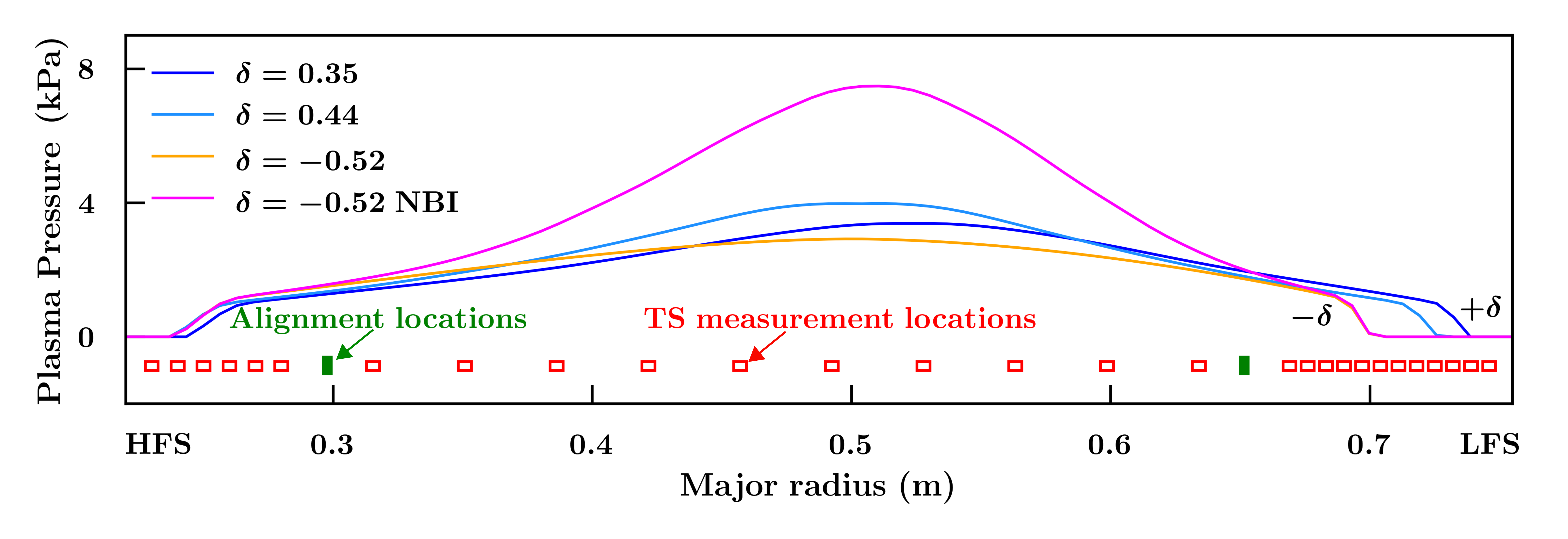}
\caption{\label{fig:pressure_prof} This image shows the location of the TS measurement (the red empty rectangles) and alignment (the green rectangles) points along the major radius of SMART. These TS locations are determined from the qualitative variation of the plasma pressure along the SMART major radius at different PT and NT values. The plasma pressure profiles were obtained by Mancini \textit{et al.}\cite{Mancini_2023} in ASTRA simulations using a GyroBohm model. }
\end{figure*}

The optimal measurement locations and spacing between adjacent scattering points are determined by considering the expected variation in radial plasma pressure profiles at different triangularity values as predicted in ASTRA simulation using a GyroBohm model \cite{Mancini_2023}. The TS measurement locations and simulation data are shown in Fig.~\ref{fig:pressure_prof}. Please note that the simulation profiles in the NT configuration are calculated by running ASTRA in predictive mode and using transport coefficients corresponding to ST plasmas with PT, as no machine has ever explored the combination of NT and low aspect ratio. Recently, more advanced simulations have been performed using the TRANSP code and the Multi-Mode Model that provide similar central values\cite{Podesta_2024, Zabala}.

Simulations suggest a narrow pedestal width, $\sim 1.2$~cm, on LFS of the magnetic axis, and a slightly wider pedestal, $\sim 2$~cm, on HFS due to the Shafranov shift. Additionally, the pedestal position on LFS shifts radially with a change in triangularity; however, this shift is not very prominent on HFS. 
The measurement points are therefore judiciously distributed to ensure dense coverage in both the LFS and the HFS regions, enabling detailed resolution of the edge pedestal in both PT and NT configurations. 
The details of the TS measurement points distribution in SMART is shown in Table~\ref{table:scattering_length}. More than a third of the measurement points are dedicated to LFS of the magnetic axis, six to the HFS pedestal, and a third are uniformly arranged in the core region. 

\begin{table}[h!]
 \begin{tabular}
 { |p{3.3cm}|p{1.2cm}|p{1.2cm}|p{1.2cm}| }
 \hline
 \textbf{Location} & \textbf{LFS} & \textbf{HFS} & \textbf{Core} \\
 \hline
 Scattering length [mm] & $5.8$ & $9.2$ & $\leq9.2$  \\
 \hline
 Separation between measurement points [mm]& $7$ & $10$ & $35.4$  \\
 \hline
 Number of measurement points & $12$ & $6$ &$10$ \\
 \hline
\end{tabular} 
\caption{This table shows the distribution of TS measurement points along the major radius of the SMART tokamak. Here the scattering length refers to the light collection length along the laser beam that determines the scattered signal strength. Twelve measurement points on LFS are dedicated to providing a spatial resolution of $7$~mm, and six on the HFS pedestal achieve $10$~mm spatial resolution. Ten measurement points are uniformly arranged in the core region to provide a spatial resolution of $35.4$~mm.   }
\label{table:scattering_length}
\end{table}

\begin{figure}
	\includegraphics[scale=0.85]{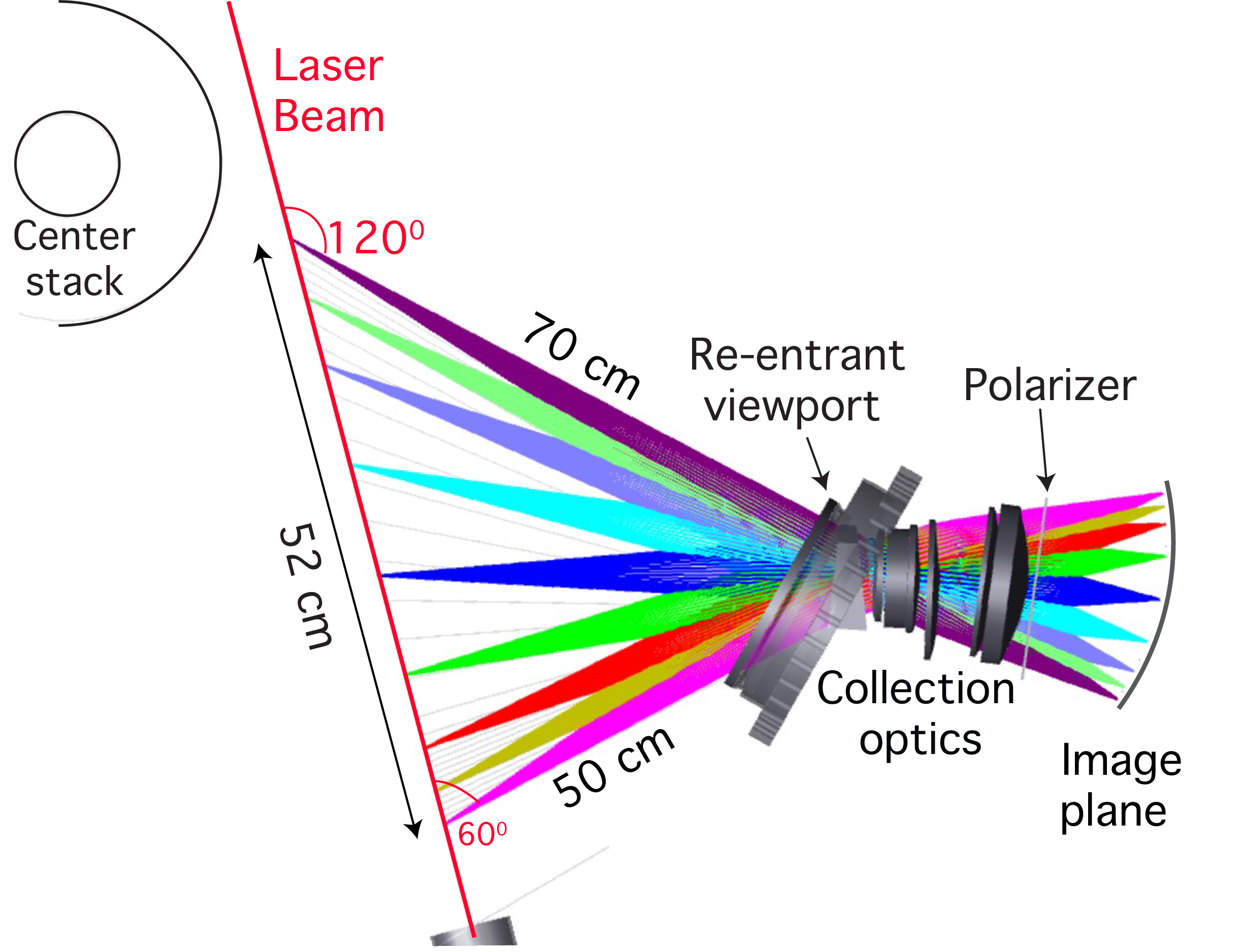}
	\caption{\label{fig:sightlines} This image shows the lines of sight of the Thomson scattering diagnostic in the mid-plane (the top view). There are 30 scattering locations, 28 of which are for measuring localized plasma parameters, and two are dedicated to monitoring the laser alignment with respect to the collection optics. }
\end{figure}

Radial coverage of $52$~cm is achieved with $28$ TS measurement points with a scattering angle ranging from $60^\circ$ to $120^\circ$, as shown in Fig.~\ref{fig:sightlines}. The wide radial coverage is made possible using a custom reentrant quartz viewport that is designed to bring the collection optics closer to the laser line, and hence the measurement points. The distance of the measurement points from the collection optics varies from $50$~cm to $70$~cm. The clear diameter of quartz in the reentrant viewport is larger than $18$~cm, enabling the use of lens elements with large aperture. There are six elements in the collection optics with their diameter varying from $9.2$~cm (for the first and smallest element) to $15.7$~cm (for the last and largest element). 
These design considerations ultimately enable us to employ fast collection optics and collect a strong light signal.

\begin{figure}
	\includegraphics[scale=0.8]{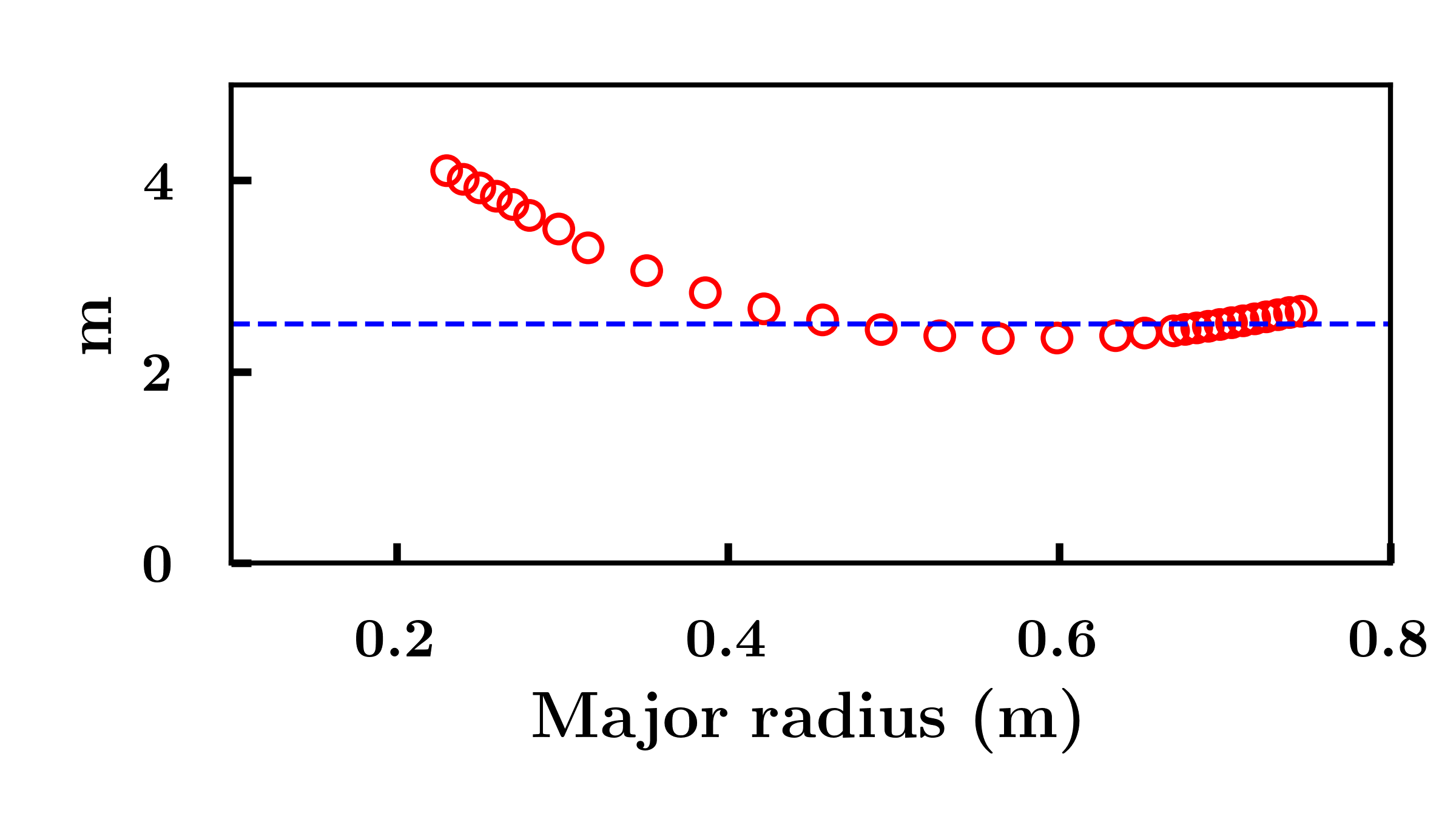}
	\caption{\label{fig:magnification} This image shows $m$ (in the red circles), the optical magnification, of the collection optics along the major radius of SMART. The blue dotted line corresponds to $m=2.5$. }
\end{figure}

Furthermore, the collection optics is designed to offer a low optical magnification, $m$, of $2.5 \times$ for the entire LFS measurement range, as shown in Fig.~\ref{fig:magnification}. 
Low $m$ ensures that the projection of the optical fiber onto the laser line does not exceed the desired scattering length of $6$~mm on LFS and hence provides us with $7$~mm  spatial resolution (i.e., the center-to-center separation between adjacent measurement points). The optical $m$ increases to $4 \times$ for the HFS pedestal and some of the core scattering locations. This makes the scattering length to vary along the laser beam from $5.8$~mm at the outboard edge to $9.2$~mm at the inboard edge.

The collection optics assembly will be housed in a dark enclosure stationed outside the vacuum vessel, and light will be collected from the midplane.  
An achromatic prism is used in front of lens elements in the collection optics to deviate its optical axis by $10^\circ$. This prevents the optical path from clashing with the nearby poloidal coils. Decentered and tilted optical elements are used to correct asymmetric aberrations arising from the achromatic prism. Using decentred and tilted elements also avoids using a shallow cylindrical optical surface on the prism, which is difficult to manufacture. The resultant optical system uses all conventional spherical lens elements, which are both easier to manufacture and assemble, while achieving the same optical performance as a more complicated system with toroidal surfaces. 
At the end of the collection optics, a wire grid polarizer will be installed to reduce randomly polarized plasma background light and collect only the polarized TS light. 

The collection optics images the laser beams onto a continuous fiber bundle holder for mounting $30$ rectangular Optical Fiber Bundles (OFB). All OFBs are made up of low-OH pure silica core/polymer cladding 210/230-micrometer diameters with their jackets removed and a numerical aperture (NA) of $0.285$.
The rectangular side of each OFB has an area of $2.3\times 1.44$~mm$^2$ and contains 66 fibers. The other end of the OFB that goes to the polychromator comprises fibers of two OFBs in a circular cross section of $3$~ mm diameter in a random distribution. The two legs of the composite OFB have different lengths, permitting time multiplexing of the data. More details are provided in the next subsection.

\subsection{Measurement of TS spectrum from collected light}
Measurement of the TS spectrum involves recording the intensity of scattered light at different wavelengths. The wavelengths needed to be measured is dependent on the $T_{\rm{e}}$ range to be measured. 
To resolve the expected $T_{\rm{e}}$ range of SMART discussed in section~\ref{Sec:Overview}, we purchased 14 polychromators from UKAEA \cite{Scannell_2008}. Each polychromator is designed to accommodate up to seven spectral channels (Ch), but is currently equipped with five optical interference filters. 
The central wavelength and bandwidth of these interference filters and the quantum efficiency of the avalanche photodiodes (APD that converts the optical signal into voltages) at the central wavelength are listed in Table~\ref{table:polys}. The additional two channels in the polychromators can be populated in the future to cover a wider $T_{\rm{e}}$ range. 

\begin{table}[h!]
 \begin{tabular}
 { |p{1.6cm}|p{1.1cm}|p{1.1cm}|p{1.1cm}|p{1.1 cm}|p{1.1 cm}|p{1 cm}|  }
 \hline
 Parameters & Ch 1 & Ch 2 & Ch 3 & Ch 4 & Ch 5 \\
 \hline
 %Central wavelength (nm) & $1060.1 - 1062.1$  & $1054.95 - 1060.45$  & $1040 - 1055$ & $995 - 1040$ & $840 - 994.5$ \\
 Central wavelength (nm) & $1061.1$ & $1057.7$ & $1047.5$ & $1017.5$ & $917$ \\
 \hline
 Bandwidth (nm) & $2$ & $5.5$ & $15$ & $45$ & $155$ \\
 \hline
 Quantum efficiency (\%) & 36.2 & 38.9 & 44.5 & 62.5 & 87.6 \\
 \hline

\end{tabular} 
\caption{This table lists the central wavelength and bandwidth (full-width at half-maxima) of five interference filters mounted onto the ordered UKAEA polychromators 
and the quantum efficiencies of the APDs at the central wavelengths. The short bandwidth filter, i.e., Ch 1, enables low $T_{\rm{e}}$ measurements.  }
\label{table:polys}
\end{table}

To optimize the use of polychromators, each polychromator will collect a scattered light signal from two scattering locations in the mid-plane using an unequal length bifurcated OFB. The length of two bifurcated bundle legs is $12$ m and $22$ m. The length difference is kept as $10$ m to introduce a time separation of $48$ ns between the signals from two scattering locations. 
This scheme enables $14$ polychromators to provide information on plasma parameters from $28$ different radial locations, and hence offers better spatial resolution and coverage than that provided by $14$ radial locations. Rotational Raman calibration will be used to determine the sensitivity of the light signal from different scattering locations in the midplane that will be used in density determination \cite{Howard_1979}. Because of the sensitive nature of the calibration on the laser alignment and the possibility of quartz window degradation overtime due to coating, the calibration will be performed both during the installation of the TS system and at the start of every major SMART campaign.

\subsection{Laser alignment monitoring }
\begin{figure}
	\includegraphics[scale=0.47]{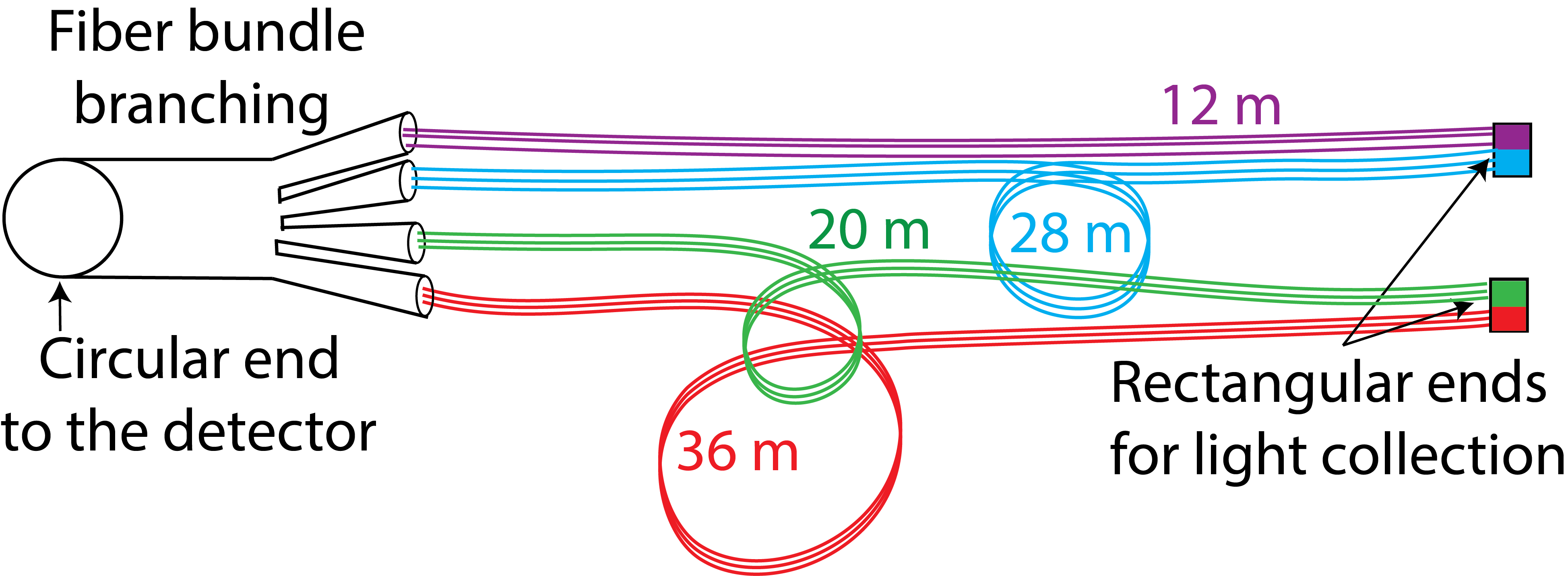}
	\caption{\label{fig:quad_fiber} This image shows a quad-OFB that is used with a one-channel \textit{polychromator} for monitoring laser alignment with respect to the collection optics. The quad-OFB has four unequal-length OFB legs to introduce a time delay between the signals from different points. }
\end{figure}

We developed a novel set up to actively monitor \textit{in-situ} laser alignment during each shot such that corrective actions can be taken in between shots if any misalignment occurs. 
In this setup, two scattering locations located in the core (the same midplane as the measurement points) are dedicated to monitoring laser alignment using a quad-OFB and a one-channel \textit{polychromator}. The quad-OFB has four unequal-length legs, as shown in Fig.~\ref{fig:quad_fiber}, to introduce time delay between signals from four legs. Each scattering location is covered by two legs of the quad-OFB arranged vertically in a rectangle, of area $1.44$~mm$\times 2.3$~mm, such that one bundle collects light signal from the top half of the laser line whereas the second collects light from the bottom half. 

The one-channel \textit{polychromator} is also purchased from UKAEA along with the other 14 polychromators. It is equipped with one interference filter with a passing wavelength band of $1040-1055$~nm (Ch 3 of Table~\ref{table:polys}). 
The radial position of the alignment fibers and passing wavelength band of the interference filter are selected such that the two measurement points are farthest apart in the core and gather a strong signal during Raman calibrations and day-to-day operations, while ensuring least contribution from Rayleigh light. Using the same detector (i.e., on APD on the polychromator) ensures that all the OFB legs, and hence the four positions (radial and vertical), record equal signal strength for a perfectly aligned laser beam with respect to the collection optics. 

\section{Summary} \label{Sec:summary}
We described the design of a multipulse Thomson scattering system to measure the electron temperature and density profiles in the mid-plane of the SMART tokamak. Special emphasis was laid on measuring the pedestal region on LFS and HFS of the magnetic axis with high spatial resolution to compare PT and NT plasma scenarios.
The diagnostic will operate with high spatial resolution, $7$~mm separation between adjacent points in the LFS pedestal region and $10$~mm in the HFS region, and a wide dynamic range, $1$~eV to $1$~keV, to resolve large gradients formed at the plasma edge and in SOL under different triangularity conditions and low aspect ratios. 
A fast collection optics enables efficient use of the fourteen polychromators available to collect the plasma information at $28$ spatial locations covering the entire plasma width along with the outer midplane SOL. In addition, the collection optics offer a low $m$ in the LFS region, allowing tight spacing between measurement points and high spatial resolution as mentioned before. A novel setup is developed to monitor \textit{in situ} laser alignment using a polychromator equipped with one interference filter and a quad-OFB at two scattering locations in the core during Raman calibrations and day-to-day operations. 
The design and order of most major parts of the SMART Thomson scattering (laser, polychromators, data acquisition, etc.) are complete. 

\vspace{-4mm}
\section*{acknowledgments}
\vspace{-4mm}
This work was primarily supported by the U.S. Department of Energy, Office of Science, Office of Fusion Energy Sciences under contract number DE-AC02-09CH11466. The United States Government retains a non-exclusive, paid-up, irrevocable, world-wide license to publish or reproduce the published form of this manuscript, or allow others to do so, for United States Government purposes. 
A part of the work has been carried out within the framework of the EUROfusion Consortium, partially funded by the European Union via the Euratom Research and Training Programme (Grant Agreement No 101052200 — EUROfusion). The Swiss contribution to this work has been funded by the Swiss State Secretariat for Education, Research and Innovation (SERI). Views and opinions expressed are however those of the author(s) only and do not necessarily reflect those of the European Union, the European Commission or SERI. Neither the European Union nor the European Commission nor SERI can be held responsible for them.

%\end{acknowledgments}
\vspace{-4mm}
\section*{Data Availability Statement}
\vspace{-2mm}
The data that support the findings of this study are available from the corresponding author upon reasonable request.

\bibliography{MKaur_TS_diagnostic}

\end{document}